\documentclass{article}%
\usepackage{amsmath}
\usepackage{amssymb}%
\usepackage{amsfonts}%
\usepackage{graphicx}

\begin{document}

\begin{center}
{\Huge Spin Singularities: Clifford Kaleidoscopes and Particle Masses }

Marcus S. Cohen

Department of Mathematical Sciences

New Mexico State University

marcus@NMSU.edu

\bigskip
\end{center}

{\large ABSTRACT}

Are particles singularities- vortex lines, tubes, or sheets in some global
ocean of dark energy? We visit the zoo of Lagrangian singularities, or
\textit{caustics }in a $spin(4,%
\mathbb{C}
)$ phase flow over compactifed Minkowsky space, $M_{\#}\equiv$ $S_{1}\times
S_{3},$ and find that their varieties and energies parallel the families and
masses of the elementary particles.

Singularities\textit{\ }are classified by tensor products of Coxeter groups
$<p,q,r>_{s}^{J},$\textit{\ }generated by reflections. The
\textit{multiplicity, }$s,$ is the number reflections needed to close a cycle
of \textit{null zigzags: }nonlinear \ resonances of $J$ \ chiral pairs of
\textit{lightlike }matter spinors with $(4-J$ $)$ \textit{Clifford mirrors}:
dyads in the remaining unperturbed vacuum pairs\textit{.} Using singular
perturbations to "peel" phase-space singularities by orders in the vacuum
intensity, we find that singular varieties with quantized mass, charge, and
spin parallel the families of leptons $(J=1)$, mesons $(J=2)$, and hadrons
$(J=3)$. Taking the symplectic 4 form, $(\psi^{I}\mathbf{d}\psi_{I})^{4}$- the
volume element in the 8- spinor phase space- as a natural Lagrangian, these
singularities turn out to have rest energies of $\ m=(s/2)^{3}m_{e},$
\textit{within a few percent of the observed particle masses}.

\subsection{Spin Space: the Vacuum as a Nonlinear Medium\bigskip}

A nonllinear mediium near a critical point typically displays
\textit{dynamical symmetry breaking}: the spontaneous formation of patterns.
Microscopic perturbations may seed a wide variety of macroscopic forms in a
nearly-homogeneous medium; for example, snowflakes, seeded by dust grains, and
subject to slightly-different histories of vapor pressure and temperature.

The interactions of critical patterns - attraction or repulsion, merger or
splitting, comes along "for free", through the same nonlinearity that created
them. Examples include hydrodynamic vortices [Newell], scroll rings in
reaction -diffusion systems [Winfree], limb-buds in morphogenesis. Neu showed
that vortices in superconductors interact via an effective electromagnetic
potential $d\vartheta^{0}:$ the far-field $u(1)$ phase differential, sourced
in quantized charges; their winding numbers, $\int d\vartheta^{0}=2\pi n$
[Neu]. Skyrme found $SO(3)$ monopoles in a "vacuum" with a quartic
nonlinearity, and showed they interacted via an effective $so(3)$ potential
[Sk]. Skyrmions were used to model hadrons by Witten, Weiss and Jackson, and
others [Witt1], [W+J], [ ],[ ].The orbit space of interacting Skyrmions was
worked out by Manton, Temple-Raston, and others. Witten used his version of
the quartic Weiss-Zumino term to model particles with quantized baryon number
and strong interactions [Wit]. This rekindled excitement in string theories,
mostly of Kaluza-Klein type, which view the "internal" degrees of freedom of
particles as compactified spacetime dimensions.

We take the opposite approach here. We view \textit{spinors }as the ultimate
reality, and spacetime as a neutral submanifold of spin space: an 8-spinor
bundle, $\Psi,$ with fiber group $spin(4,%
\mathbb{C}
).$ Particles are Lagrangian singularities in the local \textit{spinfluid
}flow, $(\Psi,\mathbf{d\Psi)\in T}^{\ast}\Psi$. Their \ regular "tails" give
effective \textit{electroweak }gauge potentials in the $PT-$antisymmetric
limit [MC1];

\begin{center}
$\Psi^{I}\mathbf{d\Psi}_{I}\mathbf{\rightarrow}g^{-1}dg=d\vartheta^{\alpha
}(x)\sigma_{\alpha}\in\lbrack u(1)\oplus su(2)]_{loc}.$
\end{center}

The $PT-$symmetric limit gives effective \textit{gravitostrong }potentials
[MC2].The antiHermitan part of their spin curvatures gives effective
$u(1)\oplus su(3)$ fields; the Hermitian part gives Clifford-algebra valued
tetrads and metric tensors [MC3]. These pull back to inertial frames in
spacetime via the \textit{spin map (see appendix)}. The gravitational field
emerges as the curvature of the dilation-boost flow of the spinfluid
[MC4].This $spin(4,%
\mathbb{C}
)$ model realizes the vision of Anandan [Anan], in which the matter fields are
the physical reality, and spacetime- with its gauge, curvature, and metric
fields, \textit{emerges} through stationerizing an action functional.

Now any admissable action must be invariant under the spin isometry group, or
\textit{Einstein Group }\textbf{E }of coordinated "external" transformations
and their "internal" (spin- space) representations. This is \textit{general
covariance. }It demands that the external and internal frames, and their
differentials, transform in complementary ways. For example, a rotation of the
external frame by $2\pi$ radians demands rotations of the internal $l$ and
$r$- chirality spin frames by $\pi.$ This suggests that the that the
\textit{matter }spinors that span the internal spin frames and the
\textit{geometric }spinors that factor the external frames are different
sections of some global spinor field that pervades all of space, including the
supports of particles. Here, it is the vacuum spinors that mediate the
interaction between the "distant masses" and localized matter spinors, giving
a mechanism for Mach's principle. Meanwhile, invariance indicates a unique
Lagrangian: the volume $[\psi^{I}\mathbf{d}\psi_{I}]^{4}$ in \textit{spin
space }$(\Psi,\mathbf{d\Psi)}$- which gives the \textit{righrt mass ratios for
the particles}.

General covariance is \textit{automatic }if our spacetime is a horizontal
local section of an 8\textbf{- }\textit{spinor bundle, }\textbf{8;}
a\textit{\ }different section for an observer in a different frame.\textbf{\ }%
We treat the spinors here as the \textit{real, physical }objects, and
spacetime vector and tensor fields as horizontal projections$.$ We call this the

{\LARGE Spin principle,(P1).} The{\LARGE \ }\textbf{4 }spinors $\psi^{I}%
\equiv\{l^{+},r^{+},l^{-},r^{-}\},$and \textbf{4 }provisionally independent
cospinors $\psi_{I},=\{r_{-},l_{-},r_{+},l_{+}\}$ , are the real physical
objects. The $\mathbf{\psi\equiv\{}\psi^{I},\psi_{I}\},$ together with their
differentials, live in \textit{spin space, }$\{\mathbf{\psi},$ $\mathbf{d\psi
}\}\in T^{\ast}\mathbf{\psi}:$ \ the space of $spin(4,%
\mathbb{C}
)$ flows.The matter fields emerge as codimension.-$J$ singularities in the
projections of $J$ chiral pairs to a horizontal base space: our spacetime, $M
$. This leaves $(4-J)$ pairs of regular \textit{vacuum spinors }to factor the
Clifford tetrads, the gauge fields, and the coupling constants.

Each spinor has spin weight $\pm\frac{1}{2},$ and conformal weight (dimension)
$\frac{1}{2}$ [Pen\}. It takes \textbf{4 }spinors to make the pseudoscalar
inner product, involving the Dirac conjugate,

\begin{center}
$\overline{\psi}_{J}\psi_{I}\equiv\psi_{J}^{T}[iq_{2}(x)]\psi_{I}\equiv
\psi^{J}\psi_{I};$ $\ \ \ i[q_{2}(x)]\equiv i[l_{1}(x)\otimes r^{1}%
(x)-l_{2}(x)\otimes r^{2}(x)]\equiv%
\begin{bmatrix}
0 & 1\\
-1 & 0
\end{bmatrix}
.$
\end{center}

Two "invisible" \textit{vacuum spinors }are hiding in the "metric spinor",
$i[q_{2}(x)],$expressed as an antisymmetric matrix (spin-1) in the
\textit{moving} spin frames. It takes \textbf{4}$\ $spinors make the
\textit{metric tensor }(spin 2),%

\begin{equation}
g_{\alpha\beta}=\frac{1}{2}[q_{\alpha}\otimes\overline{q}_{\beta}\oplus
q_{\beta}\otimes\overline{q_{\alpha}}],
\end{equation}

and \textbf{8 }spinors to make an \textbf{E-} invariant inner product,
$q^{\alpha}g_{\alpha\beta}q^{\beta}.$

It takes \textbf{4 }spinors and \textbf{4 }spinor\textbf{\ }differentials to
make an\textbf{\ E}-invariant Lagrangian 4 form; this must be a C-scalar
$(\thicksim\sigma_{0})$ to be invariently integrated. We choose the volume
element in the complexified cotangent bundle\textit{,} $\mathbf{\psi}\pm i$
$\mathbf{d\psi\in%
\mathbb{C}
(}T^{\ast}\Sigma)$ as a natural Lagrangian :

\begin{center}
$S_{g}=\frac{1}{2i}\int_{M}\left[  (\psi^{I}-i\mathbf{d}\psi^{I})\wedge
(\psi_{I}+i\mathbf{d}\psi_{I})\right]  ^{4}$
\end{center}

(sum on $I=1,2,3,4$)$.$ Here $\left[  {}\right]  _{0}$ means "the
\textit{scalar} part of a complex-Clifford $(%
\mathbb{C}
C)$ valued form". Its physical interpretation is the 4-form part of the free
product of "jets" $(\mathbf{\psi},$ $\mathbf{d\psi),}$ representing flows in
spin space. Physically, each spinor $\psi_{I}$ interacts freely with each
cospinor $\psi^{I}$ and its differential, $\mathbf{d}\psi^{I}.$ Only the $%
\mathbb{C}
C$ \textit{scalar} part of the 8-spinor tensor product can contribute to the integral.

Action $S_{g}$ is stationarized in either the $PT$-symmetric $(PT_{s})$ or
$PT$-antisymmetric $\left(  PT_{a}\right)  $ case, $\psi^{I}\mathbf{d}\psi
_{I}=\pm\mathbf{d}\psi^{I}\psi_{I}$ [SGGU]\textit{. }Because summing over an
orthonormal spin frame $\psi^{I}\equiv\{l^{+},r^{+},l^{-},r^{-}\}$ gives the
$Trace$, the Lagrangian then reduces to the 8-spinor \textit{factorization
}$[\psi^{I}\mathbf{d}\psi_{I}]^{4}$ of the \textit{Maurer-Cartan (M.C.) 4
form}, $Tr[g^{-1}\mathbf{d}g]^{4}$ for $Spin(4,%
\mathbb{C}
)=%
\mathbb{C}
(U(1)\oplus SU(2)).$ The integral of the unitery part $U(1)\oplus SU(2)$ is
quantized over compactified Minkowsky space, $M_{\#}\equiv$ $S_{1}\times
S_{3}:$

\begin{center}
$\int_{M_{\#}}Tr[g^{-1}\mathbf{d}g]^{4}=16\pi^{3}N$ .

This provides a \textit{bound }for the stationery \textbf{8}-spinor action,
which is saturated in either the $PTa$ \ or $PTs$ limit,

$\psi^{J}\rightarrow\mp\psi_{J}^{T}[iq_{2}(x)]=\mp\psi_{J}^{T}%
\begin{bmatrix}
0 & 1\\
-1 & 0
\end{bmatrix}
,$
\end{center}

in the \textit{moving} spin frame [SGGU] . In either case, the stationery
action becomes

\begin{center}
$S_{g}\equiv\int_{M_{\#}}\mathcal{L}_{g}\rightarrow\int_{M_{\#}}\left[
(\psi^{1}\mathbf{d}\psi_{1})\wedge(\psi^{2}\mathbf{d}\psi_{2})\wedge(\psi
^{3}\mathbf{d}\psi_{3})\wedge(\psi^{4}\mathbf{d}\psi_{4})\right]  _{0}%
\equiv\int_{M_{\#}}[\psi^{I}\mathbf{d}\psi_{I}]_{0}^{4}=16\pi^{3}N.$
\end{center}

\ Stationarizing $S_{g}$ cuts out a minimal surface in spin space.

Note that invariance of $S_{g}$ under coupled external and internal \textbf{E
}transformations is \textit{automatic }if the matter spinors, $\psi,$ are
envelope modulations riding on a background o\textit{f vacuum spinors}:
$\psi(x)\equiv l^{A}\mathbf{(}x\mathbf{)\psi}_{A}(x)$ (sum on $A=(1,2)).$ The
differentials of the moving spin frames, $\ell^{A\mathbf{\ \ }}(x)$ then
appear as \textit{gauge fields} in the covariant derivative:%

\begin{equation}%
\begin{array}
[c]{c}%
\mathbf{d}\psi_{A}\equiv\mathbf{d}\left(  \ell^{A\mathbf{\ \ }}{}\psi
_{A}\right)  =l^{A}\mathbf{d}\psi_{A}+\mathbf{d}l^{A}\psi_{A}=l^{A}\left(
\partial_{\alpha}+\Omega_{\alpha}\right)  \psi_{A}e^{\alpha}\\
\equiv l^{A}\nabla_{a}\psi_{A}e^{\alpha}.
\end{array}
\end{equation}

Serenditously, although $\mathcal{L}_{g}$\ is a natural 4 form with no
coupling constants, the $(4-J)$ \textit{unperturbed vacuum spinors
}make\textit{\ effective coupling constants }of the proper dimensions- and the
right magnitudes- to couple $J$ matter or gauge -field envelopes [$\chi U$ ]!
In the regular,\textit{\ geometrical-optics} regime$,$ $S_{g}$ yields the
proper effective actions for electroweak (PTa, or charge-separated) and
gravitostrong (PTs , or neutral) fields, $\mathbf{dd}\psi=\mathbf{\kappa\psi
.}$[ ] Here it agrees with Witten's "\emph{Weiss-Zumino} 4 form," action,
$\int Tr(g^{-1}dg)^{\wedge4}g$ , which is quantized over the boundary,
$\gamma_{4}\sim\partial B_{5},$ of a 5-manifold [Witten1], [Witten2]. We could
find $B_{5}\sim\mathbb{C}\times\mathbb{S}_{3}$ embedded in our position-world
velocity phase space $\ z^{\alpha}=x^{\alpha}+y^{\alpha}\in$ $\mathbb{C}_{4,}$
with \textit{complex time} coordinate $z^{o}=t+iT$; provided all particles
were at rest, $(y_{j}=0).$ In the \textit{spinfluid regime, }there is a unique
flow world velocity, $y^{\alpha}(x^{\alpha}),$ at each spacetime point, that
varies continuously from point to point $x^{\alpha}\in M.$

\bigskip Geometrical optics breaks down on \textit{boundary caustics,}
$\gamma_{4-J}$ $\in$ $\partial B_{5-J}$, where, the spin map $S:$
$TM\rightarrow T\Sigma$ becomes \textit{singular, }and acquires a
$J-$dimensional kernel. The domains these caustics enclose are
\textit{branched covers,} with $J$ extra\textit{\ }bispinor sheets in spin
space over each spacetime point These accommodate the wave functions of
$J-$bispinor particles.

Caustics arise in optics, hydrodynamics, chemical reactions, acoustics, etc.
as loci of partial focusing, or shock fronts [Arnold]. Joe Keller, Alan Newell
[Newell], and others have used a powerful tool to look inside these apparent
singularities: \textit{singular perturbation theory} or multiscaling; defining
a short spacetime scale inside the shock, and matching the inner solution to
the outer one on the shock boundaries. We apply it to give a system of coupled
envelope-modulation equations [Newell] to nonlinear waves in the 8 spinor
medium: the \textit{spinfluid, }and find\textit{\ }that their caustics are the
\textit{elementary particles}. We outline the results below; details of the
calculations appear in Part III [M.C. 4].We previously showed how a
homogeneous background of \textit{vacuum spinors} could produce this chiral
cross coupling : Mach's principle for bispinor particles[I.M.,U]. We derived
Einstein's equations, using integration by parts to match the effective
actions inside and outside the worldtube boundary: the inertial mass and the
\textit{gravitational mass.}

What we do here is

1) show how nonlinear interactions with the vacuum spinors "fold up"
\textit{lightlike }spin rays inside a\textit{\ timelike }worldtube, $B_{4}$.

2) Identify these folding patterns with the varieties of
\textit{singularities- and the families of elementary particles.}

3) Classify these by the \textit{Coxeter groups of multiplicity-}$s:$ the
$s$-fold covers of the Weyl reflection groups. .

4) Derive the formula $m=(s/2)^{3}m_{e}$ for the \textit{masses }of the
corresponding particles in terms of their \textit{Coxeter numbers, }%
$s$\textit{.}

\section{Singularities and Stratification}

\bigskip In the g\textit{eometrical-optics (g.o.) }regime, $D^{0},$
\textit{regular} phase flows are created by nonsingular active-local (perhaps,
path-dependent) Einstein transformations, $\left(  L(x),R(x)\right)  $\ $\in
E_{A}$, written as $GL(2,C)$ matrices in the complexified Clifford-algebra $%
\mathbb{C}
C(R_{+}\times S_{3}).$ These act on a fiducial spin frame of \textit{vacuum
spinors}, $\left(  \hat{\ell},\hat{r}\right)  $, written column wise and row
wise respectively$:$%

\begin{equation}%
\begin{array}
[c]{c}%
\ell\left(  x\right)  =\hat{\ell}\exp\left[  \frac{i}{2}\zeta_{L}^{\alpha
}(x)q_{\alpha}\right]  \equiv\hat{\ell}L\left(  x\right)  ;\\
r\left(  x\right)  =\exp\frac{i}{2}\zeta_{R}^{\alpha}(x)\overline{q}_{\alpha
}]\hat{r}\equiv R\left(  x\right)  \hat{r},
\end{array}
\end{equation}

In the regular$,PT-$antisymmetric ($PTa$) case, $R\left(  x\right)
=L^{-1}\left(  x\right)  ,$ multiplying a spinor by the differential of the PT
opposed spinor gives effective \textit{spin connections, or} \textit{vector
potentials}$:$ Clifford- algebra-valued 1 forms,%

\begin{equation}%
\begin{array}
[c]{c}%
\Omega_{L}\equiv\ell^{-1}\mathbf{d}\ell\left(  x\right)  =\mathbf{d}\zeta
_{L}=\left[  \partial_{\alpha}\zeta_{L}^{a}\right]  (x)q_{a}e^{\alpha}%
\text{,}\\
\Omega^{R}\equiv\left(  \mathbf{d}r\right)  r^{-1}\left(  x\right)
=\mathbf{d}\zeta_{R}=\left[  \partial_{\alpha}\zeta_{R}^{a}\right]
(x))\overline{q}_{a}e^{\alpha}\text{.}%
\end{array}
\end{equation}

However, even for a regular initial distribution of spinor fields,
codimension- $J=\left(  1,2,3,4\right)  $ \textit{phase singularities }%
$\gamma\left(  4-J\right)  $ will form, shift, merge, annihilate, and
recombine, like the projections of folds in a sheet to the bed. In addition to
the \textit{regular stratum, }$\gamma^{o},$where the projection $\pi$ from the
Lagrangian submanifold of spin space solutions to the position-world velocity
phase space,%

\[
\pi:(\psi_{I}+i\mathbf{d}\psi_{I})\in%
\mathbb{C}
T^{\ast}\Sigma\rightarrow x^{\alpha}+iy^{\alpha}\in%
\mathbb{C}
T^{\ast}M,
\]
is 1 to 1, there will be codimension-$J=(1,2.3,4)$ singular strata:
\textit{branched covers,} $D^{J},$ where $\pi$ is $\ J+1$ to $1.$ Like the
crisscrossing rays inside a kaleidoscope, there are $J+1$ world-velocity
sheets, $y^{\alpha},$ over each spacetime point, $\ x^{\alpha}\in B_{4-J}$,
inside the support,\textit{\ }$B_{4-J},$ of a $J-$bispinor particle$.$ Each
support is bounded by\ loci of partial focusing,\emph{\ }boundary
caustics\emph{, }$\gamma_{4-J}$\emph{\ }$\subset\partial B_{5-J}$ : \ folds,
cusps, tucks, swallow-tails and knots, where spin rays $\psi^{I}\mathbf{d}%
\psi_{I}$ = $\mathbf{d}\zeta_{I}$ branch or converge [ref Arnold]. Each
$(4-J)$ \textit{brane,} $B_{4-J},$ carries a $J-$form matter current, $\ast
J,$\textit{dual} to the Clifford volume element contributed by the
$(4-J)$\textit{\ vacuum }pairs\textit{.} We call this complex of$\ $branes and
currents the Spin (4,$%
\mathbb{C}
$) \textit{complex}, or \textit{spinfoam}.\qquad\qquad\qquad\qquad\qquad
\qquad\qquad

A 3- dimensional example is a foam of soap bubbles, with the regular stratum,
$B_{3}=\ast D^{o}$ (the volumes), and singular strata, $\gamma_{2}%
\subset\partial B_{3},\gamma_{1}\subset\partial B_{2},\gamma_{0}%
\subset\partial B_{1}:$ the surfaces, edges, and vertices. Each stratum,
$D^{J},$carries a $\ J-$ form current: density in volumes $\gamma_{3}$
pressure on surfaces $\gamma_{2},$ tension in line segments $\gamma_{1}$, and
force on nodes $\gamma_{0}.$

A codimension- J \textit{bifurcation} occurs at the critical point, $x_{c}%
\in\gamma_{4-J},$where the rank of the Jacobian matrix, $\ [\mathbf{d\zeta
}\boldsymbol{]}(x_{c})\boldsymbol{\equiv}$ $[\partial_{\alpha}\zeta_{\beta
}](x_{c}),$ drops by J. Here, $J+1$ phase differentials become linearly
dependent, to span only a $\left(  4-J\right)  $-dimensional subspace. If the
Hessian, $[\mathbf{d}^{2}\zeta](x_{c})$, is singular there too, $\left\vert
\partial_{\alpha}\partial_{\beta}\zeta\right\vert (x_{c})=0$,\textbf{\ }%
$x_{c}$ is a\textit{\ degenerate critical point: }a\textit{\ }caustic, where
rays $\mathbf{d\zeta}_{I}$\ merge or split, and there is a change in the
topology of the orbits..\qquad

This is \textit{dynamical} \emph{symmetry breaking. }One tool to detect it is
the\qquad\qquad\qquad\qquad\qquad\qquad\qquad\qquad\qquad\qquad\qquad
\qquad\qquad

$\mathbf{Equivariant\ Branching\ lemma}$ (Michel's "theorem"): If the isotropy
subgroup, $H\subset E,$ that fixes a solution $\Psi_{{\Large c}}$ contains
just a single copy of the \emph{identity} representation, then $\Psi_{C}$ is a
possible direction for dynamical-symmetry breaking ref. [Sattinger].

Some corollaries are

\begin{enumerate}
\item the branched covers and boundary caustics \emph{stratify} the base
space, $\mathbb{M}$, into orbits of $E$-group actions into \emph{isotropy
subgroups,} $H:$%
\[
\mathbb{M}=\bigcup_{J=0}^{4}B_{4-J}\oplus\gamma_{4-J}.
\]

\item Generically, as you cross a \textit{boundary\ caustic} $\gamma
_{3-J}\equiv\partial B_{4-J},$ where \
$\vert$%
$dd\zeta|=0,\ker d\zeta$ picks up generators one at a time

\item The boundary of each stratum consists of singular loci belonging to the
next higher stratum,\emph{\ }except where two caustics intersect. Here, their
co-dimensions add:%
\[
\gamma_{4-J}\cap\gamma_{4-K}=\gamma_{4-M}:M=J+K\text{.}%
\]

\end{enumerate}

A bifurcating pattern is a new identity representation if it is steady state;
a stationery flow is an identity representation on the position-velocity phase
space. What is flowing in this picture is the 8-spinor vacuum; the
\textit{spinfluid.} What we\textit{\ are }looking for are patterns that
bifurcate \textit{locally} as it expands, like snowflakes from a saturated cloud.

Localization involves gluing conditions for splicing a compact "bubble",
$\Omega^{J}\equiv(\psi^{I}d\psi_{I})^{J}$\ of $J$ matter-spinor pairs into the
vacuum distribution. These give constraints on\ their integrals; not only on
spacelike surfaces, but on initial and final temporal boundaries. As the neck
of the J-tube $\gamma_{4-J}$ joining the matter bubble and the vacuum
background expands from a single point, the matter wave functions must
\textit{match} the vacuum spinors there. This gives quantized
\textit{topological charges} [Taubs] , [Uhlenbeck]: integral periods for J-
form matter currents over compactified \textit{spacetime} cycles; e.g. Bohr orbits,

\begin{center}
$\int_{\gamma_{1}}pdq-Edt=(n+\frac{1}{2})\hslash.$

The\textit{\ spacetime }holonomy operator in the perturbed vacuum surrounding
a charge is the electric field, $E_{r}\equiv F_{or}$ .

For wavefunctions of \textit{definite spin }(i.e. with only upper or a lower
complex coordinates)\textit{, }Milnor's \textbf{Fibration Theorem}%
\textit{\ }[Milnor] guarantees a complete set of \ $(4-J)$
\ \textit{parallelizable}\ fiber coordinates bridging the perturbed vacuum
between singular loci: the integral curves of \ the
\textit{vacuum\ spin\ forms}\emph{,}$\emph{\ }\widehat{\Omega}^{4-J}$ (Table I).
\end{center}

\newpage

\begin{center}
{\Large Table I: the\ vacuum\ spin forms,}

Assuming the vacuum spinors all have the same amplitude, $k^{\frac{1}{2}},$%

\[%
\begin{array}
[c]{c}%
\hat{\Omega}=\pm\left(  \frac{ik}{2a}\right)  q_{\alpha}e^{\alpha}\\
\hat{\Omega}^{2}=\left(  \frac{ik}{2a_{\#}}\right)  ^{2}q_{\ell}\left[
\epsilon_{jk}^{\;\ell}e^{j}\wedge e^{k}\pm e^{0}\wedge e^{\ell}\right] \\
\hat{\Omega}^{3}=\pm\left(  \frac{ik}{2a_{\#}}\right)  ^{3}q_{\ell}%
\epsilon_{jk}^{\;\ell}e^{j}\wedge e^{k}\wedge e^{0}\pm i\epsilon_{jk\ell}%
q_{0}e^{j}\wedge e^{k}\wedge e^{\ell}\\
\hat{\Omega}^{4}=\left(  \frac{ik}{2a_{\#}}\right)  ^{4}q_{0}\left[
\epsilon_{\alpha\beta\gamma\delta}e^{\alpha}\wedge e^{\beta}\wedge e^{\gamma
}\wedge e^{\delta}\right]  =\frac{3}{2}\left(  \frac{k^{4}}{a_{\#}^{4}%
}\right)  \mathbf{d}^{4}V,
\end{array}
\]

\bigskip
\end{center}

The constraint that the Lagrangian density must be a C scalar assures that
only the parts of $\hat{\Omega}^{4-J}$ both \textit{Clifford }and
\textit{Hodge} \emph{dual} to the matter forms, $\tilde{\Omega}^{J}%
\equiv\left(  \psi^{I}\mathbf{d}\psi_{I}\right)  ^{J},$ to contribute to the
action. These make the Clifford line, surface, and volume elements that
multiply $\tilde{\Omega}^{J}\ $to fill out the\textit{\ }\textbf{E-}invariant
(C-scalar) $4$\emph{\ -}volume element,

\begin{center}%
\[
\left\vert (\mathbf{d}\zeta)^{4}\right\vert \sigma_{o}e^{0}\wedge e^{1}\wedge
e^{2}\wedge e^{3}:\ \thicksim\gamma^{4}(dx)^{4}.
\]

\end{center}

Any C-dual contribution to $S_{g}$ must therefore be Hodge dual, as well,
effectively quantizing $\tilde{\Omega}^{J}$ against \emph{dual}
(perpendicular) cycles, $\gamma_{4-J}$ , as well as over cycles $\gamma_{J} $
\ (e.g. \ quantization of electric, flux, $F_{or}$ , over $S_{2}(\theta,\phi)$
[M.C. 2]).

These topological charges remain constant with cosmic expansion, while the
vacuum spin forms\emph{,} $\hat{\Omega}^{4-J}$ (table I) give a factor of
$\ k^{4-J}\sim\gamma^{J-4}$ \ to the action contributed by the $D^{J}$
stratum. \ Integrating in the comoving frame, $E^{\alpha}=\gamma e^{\alpha} $
results in a net action polynomial in the scale factor, $\gamma:$ the
\textit{effective potential},%

\begin{equation}%
\begin{array}
[c]{c}%
V(\mathbf{n,\gamma)}=\int_{D_{0}}\hat{\Omega}^{4}+\gamma\int_{D_{1}}%
\hat{\Omega}^{3}\wedge\left(  \psi^{I}\mathbf{d}\psi_{I}\right)  +\gamma
^{2}\int_{D_{2}}\hat{\Omega}^{2}\wedge\left(  \psi^{I}\mathbf{d}\psi
_{I}\right)  ^{2}\\
+\gamma^{3}\int_{D_{3}}\widehat{\Omega}\wedge\left(  \psi^{I}\mathbf{d}%
\psi_{I}\right)  ^{3}+\gamma^{4}\int_{D_{4}}\left(  \psi^{I}\mathbf{d}\psi
_{I}\right)  ^{4}=16\pi^{3}\left[  n_{0}+n_{1}\gamma+n_{2}\gamma^{2}%
+n_{3}\gamma^{3}+n_{4}\gamma^{4}\right]  ,
\end{array}
\end{equation}

where $n_{J}$ is the population of the $Jth$ stratum [M.C..3] .

The polynomial $V(\mathbf{n,}\gamma)$ can \textit{mimic the effect of the
Higgs field }by mixing positive-definite quadratic couplings in $\gamma^{2}$
with \ negative-definite quartic ones in $-$ $\gamma^{4},$ to create a
"Mexican hat" potential. But, unlike standard Q. F. T., the lepton, meson,
hadron and atomic masses appear in a\textit{\ }4-term sequence, at
$O(\gamma,\gamma^{2},\gamma^{3,}\gamma_{4}),$ respectively.

The $\hat{\Omega}^{3}$ term contributes the 3 -volume element in \textit{spin
space }to\ the Noether charge under complex-time $(z^{0}\equiv t+iT)$
translation, which includes the Jacobean determinant of the $3$-space block of
spin map, $S:$%

\[
|\left(  \mathbf{d\zeta}\right)  |^{3}\thicksim s^{3}e^{1}\wedge e^{2}\wedge
e^{3}.
\]

This gives quantization of both \textit{mass }and\textit{\ charge}:%

\begin{equation}
\int_{B_{3}}[(\partial_{t}\theta^{0})-i(\partial_{r}\theta^{0})]e^{1}\wedge
e^{2}\wedge e^{3}=M+iQ.
\end{equation}
It is the\textit{\ vacuum spinors, } hiding the Clifford 3-volume element
$\hat{\Omega}^{3},$ that endow frequency, $\omega\equiv(d_{t}\theta^{0}),$
with\textit{\ mass:\ Mach's principle} in action. Both $m$ and $q$ come in
integral units: particle numbers.

Continuity of the Gluing map [Taubs] says that the matter spinors localized
inside the compact world tube $B_{4}$ must match the vacuum distribution on
its boundary, $\gamma_{3}\equiv\partial B_{4\text{ }}.$ However, as you pass
along a curve $x\in\gamma_{1}$ through a degenerate codimension- $J=(1,2,3,4)$
boundary singularity, $x_{c}\in$ $\gamma_{4-J}\subset\partial B_{5-J},$ both
the Jacobean and the Hessian determinants vanish, and the rank of the spin map
drops by $J$:%

\begin{equation}
S\equiv\left[  \partial_{\alpha}\zeta^{\beta}\right]  \left(  x_{c}\right)
\ q_{\beta}e^{\alpha}:\left\vert \partial\zeta\right\vert \left(
x_{c}\right)  =\left\vert \partial^{2}\zeta\right\vert (x_{c})=0\Rightarrow
\ r(x_{c})=(4-J).
\end{equation}

A point inside\ $\gamma_{4-J}$ \ acquires $2J$ new preimages in the projection
$\pi:L\rightarrow M$\ \ from the Lagrangian submanifold in spin space to
spacetime. [Taubs] .

To look inside these singular loci, we use \textit{singular perturbation
theory; }what Don Cohen calls \textit{"two timing and double crossing"}.
Following Joe Keller, Alan Newell [Newell], and others, we definie a short
spacetime scale, $x=\gamma X$ inside the shock front, and match the inner
solution to the outer one on the shock boundaries. We apply it here to
caustics in the \textit{spinfluid.} We outline the results here; details of
calculations appear in Part III [M.C.4].

First, we express each spinor field as a vacuum field, $\varphi_{I}%
\equiv\left(  \hat{\ell}_{\pm},\hat{r}_{\pm}\right)  $ of amplitude
$k^{\frac{1}{2}}\thicksim\gamma^{-\frac{1}{2}},$ plus an envelope modulation:%

\begin{align}
\ell_{I}(x,X)  & =k^{\frac{1}{2}}\hat{\ell}_{\pm}(X)+\psi_{L}^{\pm
}(x)\ =\gamma^{-\frac{1}{2}}\hat{\ell}_{\pm}(X)+{\Huge \ }\psi_{L\pm}(x)\ ;\\
r_{\pm}\left(  x,X\right)   & =\gamma^{-\frac{1}{2}}\left(  X\right)
+\psi_{R\pm}.\nonumber
\end{align}

In inflated regimes, like ours, $\gamma\gg1.$ In superdense regimes,
$\ \gamma$ $\ll1;$ the matter spinors are ripples riding on the vacuum: a deep
ocean of \emph{dark energy}. Since solutions are either symmetric or
antisymmetric about the critical radius, $a=a_{\#};$ $\gamma=1$, we can
consider either case, and cover both [M.C.1]. Inserting \textit{ansatz
}$\mathbf{(}17\mathbf{),}$we obtain effective Lagrangians, $\mathcal{L}^{J}%
$\emph{,} in which $(4-J)$ vacuum pairs couple $J$ matter pairs. Varying with
respect to $\hat{\ell}_{\pm}$ or $\hat{r}_{\pm}$ gives the \emph{massless
Dirac equations}. These say that the vacuum spinors are
\textit{Clifford-analytic} and conjugate-analytic respectively:%

\begin{equation}%
\begin{array}
[c]{c}%
\overline{D}\hat{l}_{\pm}\equiv q^{\alpha}\left(  \partial_{\alpha}%
+\hat{\Omega}_{a}^{R}\right)  \hat{l}_{\pm}\left(  X\right)  =O\\
D\hat{r}_{\pm}\equiv q^{\alpha}\left(  \partial_{\alpha}+\hat{\Omega}_{a}%
^{L}\right)  \hat{r}_{\pm}\left(  X\right)  =O.
\end{array}
\end{equation}

\emph{Covariantly constant} (freely-falling) solutions, $\left(
\partial_{\alpha}+\Omega_{\alpha}\right)  \left(  \hat{l}_{\pm},\hat{r}_{\pm
}\right)  =0$ define \emph{inertial spin frames}$.$ On $M_{\#}\equiv
S_{l}xS_{3}(a_{\#}),$%

\begin{equation}%
\begin{array}
[c]{cc}%
\hat{\ell}_{\pm}\left(  X\right)  =\mathbf{\hat{\ell}}_{\pm}\left(  0\right)
\exp(\frac{i}{2a_{\#}}X^{\alpha}\sigma_{\alpha}^{\pm}); & \hat{r}_{\pm}\left(
X\right)  =\hat{r}_{\pm}\left(  0\right)  \exp(\frac{i}{2a_{\#}}X^{\beta
}\overline{\sigma}_{\beta}^{\pm});\\
\hat{\Omega}_{\pm}^{L}=\frac{i}{2a_{\#}}\sigma_{\alpha}^{\pm}e^{\alpha}, &
\hat{\Omega}_{\pm}^{R}=\frac{i}{2a_{\#}}\overline{\sigma}_{\beta}^{\pm
}e^{\beta}.
\end{array}
\end{equation}

For a given scale factor, $\gamma,$ the vacuum action is extremized when the
inertial spinors span a\textit{\ hypercube} in spin space.

Neutral combinations of vacuum spinors could be called "cosmological
neutrinos", $\nu_{l}=(\widehat{l}_{+}\oplus\widehat{r}_{-});v_{r}=(\widehat
{l}_{-}\oplus\widehat{r}_{+}).$ More generally, \emph{left} and \emph{right}
\emph{chirality} moving spin frames, $\ell_{\pm}$ and $r_{\pm}$, are given
by\emph{\ path-dependent,} \emph{active-local\ }($E_{A}$)
\emph{transformations} on the inertial spinors [M.C. 1], [M.C. 2], [M,C. 3].
These vary on the cosmic scale, so $\gamma$ beats of the logic clock, $\Delta
X^{0}=\gamma,$ elapse for each beat, $\Delta x^{0}$ $=1,$ of the local clock.%

\begin{equation}%
\begin{array}
[c]{c}%
\ell\left(  X,x\right)  \equiv\hat{\ell}_{\pm}\left(  X\right)  L^{\pm}\left(
x\right)  \text{;}\\
r\left(  X,x\right)  =\hat{r}_{\pm}\left(  X\right)  \bar{R}^{\pm}\left(
x\right)  .
\end{array}
\end{equation}

At $O(\gamma),$we obtain the massive Dirac system as our coupled-envelope
equations. Dirac mass - chiral cross coupling - appears via a \emph{spin}
$\left(  4,\mathbb{C}\right)  $ \emph{resonance}; the 8-spinor analog of
4-wave mixing in nonlinear optics [M.C. 5] .

To contribute a $C$ scalar 4 form $\sigma_{0}e^{0}\wedge e^{1}\wedge
e^{2}\wedge e^{3}$ to the action integral, a chiral pair of matter spinors
must find 3 \textit{other} pairs of vacuum spinors whose product meets the
\emph{Bragg }$\emph{(solvability)}$\emph{\ conditions; the massive Dirac
equations,}

.%
\begin{equation}%
\begin{array}
[c]{c}%
\overline{D}\psi_{I}^{L}\equiv q^{\alpha}\left(  \partial_{\alpha}%
+\tilde{\Omega}_{a}^{L}\right)  \psi_{I}^{L}\left(  X\right)  =[2a_{\#}%
]^{-1}\psi_{I}^{R}\\
D\psi_{I}^{R}\equiv q^{\alpha}\left(  \partial_{\alpha}+\tilde{\Omega}_{a}%
^{R}\right)  \psi_{I}^{R}\left(  X\right)  =[2a_{\#}]^{-1}\psi_{I}^{L}.
\end{array}
\end{equation}

The electron mass-the inverse of the critical diameter, $2a_{\#}$- comes from
the product of the 3 unbroken vacuum pairs; the $\hat{\Omega}^{3}$ in Table l.
If the vacuum spinors have different amplitudes, the scalar mass term is
replaced by the term $\psi^{I}[\hat{\Omega}^{3}]_{I}^{J}\psi_{J},$ in
the\textit{\ lepton mass matrix}$.$ This is a rank- 2 tensor product of the 6
remaining vacuum spinors C dual to $(\psi^{I},\psi_{J});$ the ones needed to
make the $\ $C-scalar $(\sigma_{0})$ term, at $O(\gamma):$ $\psi^{I}%
[\hat{\Omega}^{3}]_{I}^{J}\psi_{J}\in\langle2,\widehat{\mathbf{6}}\rangle\in
L^{1}.$

\bigskip At $O(\gamma^{2}),$integration by parts gives wave equations in
$(\overline{D}D+D\overline{D})\equiv\Delta:$ Klein -Gordon (spin l or 0)
equations, sourced in the current $3$ form, $J$, with charge quantized over 3 -cycles:%

\begin{equation}
(\Delta+\hat{\Omega}^{2})\widetilde{\Omega}=J;\quad\int_{B_{3}}J=\int_{B_{3}%
}[\partial_{o}\widehat{\theta}_{o}]\mathbf{d\zeta}^{l}\wedge\mathbf{d\zeta
}^{2}\wedge\mathbf{d\zeta}^{3}=8\pi^{2}[2a_{\#}]^{-1}B.
\end{equation}

\textit{\ }Again, the mass term, $\hat{\Omega}^{2},$ comes from the vacuum
energy. For photons, $B=0.$

At $O(\gamma^{3}),$the principal part is a system of 3 Euler equations,
coupling each quark current, $Q_{I},$ to the 2 others through the vacuum
spinors:%
\[%
\begin{array}
[c]{cc}%
\qquad Q_{I}\in\lbrack l\otimes_{I}r]l;\ Q^{I}\in r[r\otimes^{I}l]: &
\qquad\overline{D}Q_{I}\thicksim\lbrack2a_{\#}]^{-1}T_{KI}^{J}Q_{J}Q^{K}.\\
& \qquad DQ^{I}\thicksim\lbrack2a_{\#}]^{-1}T_{J}^{IK}Q^{J}Q_{K}%
\end{array}
\]

$T$ is the rank -3 "moment of inertia" tensor, with eigenvalues $I\equiv
(\ p,q,r).$ Orbits lie on invariant tori or ellipsoids, and \textit{close for
integer ratios} $(\ p/q/r),$ with a frequency that is a common multiple,
$\ s=CM(\ p,q,r)$. Pythagoras would like this; it is the condition for a
harmonious 3-note chord.

At $O(\gamma^{4}),$ we obtain a class of exact solutions we call $Spin(4,C)$
vortices; "vortex atoms" with dense nuclei of matter currents flowing in the
$+T$ direction, outward from the big bang, and diffuse shells of returning
currents, with charges $+Z$ \ and $-Z$, respectively [M.C.5]. Kelvin would
like this.

Behind all this algebraic structure lives a simple physical picture :
\textit{each chiral pair, } $q_{I}\equiv(l\otimes_{I}r),$ \textit{acts as}
\textit{a mirror for the other 3 chiral pairs, }bootstrapping from noise a
resonant $s$- cycle.

\bigskip

\section{Reflection Varieties and their Masses}

The Dirac operator, $D\equiv\sigma^{\alpha}\partial_{\alpha}:e^{\beta
}\hookleftarrow q^{\beta}$, assigns a spacetime differential to an
infinitesimal displacements in the Clifford algebra [ BDS]; [G- M].

But on \textit{boundary} \textit{caustics,} $\gamma_{3-J}\subset D$ $B_{4-J} $
the spin map, $S^{\ast}=$ $\mathbf{d}\zeta=[\partial_{\alpha}\zeta^{\beta
}]\sigma_{\beta}e^{\alpha},\ $\newline becomes singular, with rank (4 -J).
Here, some steps in internal phase no longer pull back to spacetime
increments. Meanwhile, inside there are $s$ bispinor sheets \textit{for each
spin direction }in the spin bundle over the particle support, $B_{4-J}$ .

For a volume element, $e^{4}$, to contribute to the action, the product of the
4 reflection operators in it must be a scalar. Physically, each cycle of 4
"interference gratings" $\ell\otimes r$ ---\emph{including} the curved
gratings involving \emph{matter spinors}, must close to form a
\emph{resonator}, with a net loop transfer function proportional to the
identity, $\sigma_{o}.$This closure constraint admits only a few sets of
integers $\{p,q,r,s\}$ characterizing the possible symmetry groups of singular
loci and isophase contours for particle wave functions: the\textit{\ Coxeter
groups, }$R_{s},$ [Coxeter] with their invariant polynomials in 4 complex
variables, the \textit{Breiskorn varieties }[Milnor] :%

\begin{equation}
R_{s}=\left\langle p,q,r\right\rangle _{s};\ \ \ \ \ \ \ \ B(x,y,z,T)\equiv
(x^{p}+y^{q}+z^{r}+T^{-s})=const.
\end{equation}

These are \emph{fibred knots.} For example, $\left\langle 2,2,3\right\rangle
_{3}\cap\mathbb{S}_{3}\left(  1\right)  $ is a trefoil knot, with its isophase
fibers, $\ln f\left(  x,y,z\right)  =const$, wrapping 3 times vertically
around the singular filament, $\ f\left(  x,y,z\right)  \equiv x^{2}%
+y^{2}+z^{3}=0,$ over 2 horizontal circuits.

The isophase contours over position $\ x^{\alpha}\in M$ seem to cross in the
projections,
\[
\Pi:x^{\alpha}+iy^{\alpha}\longrightarrow x^{\alpha},\text{\qquad}%
\zeta^{\alpha}=\theta^{\alpha}+i\varphi^{\alpha}\longrightarrow\theta^{\alpha
}\text{,}%
\]
from phase space to spacetime, like the crisscrossing rays in a 3D
kaleidoscope. These apparent crossings are resolved by lifting via $\Pi^{-1}
$: i.e. by separating overlapping C-algebra valued wave vectors in the
"quiver" of \ spin waves over $x^{\alpha}.$

It turns out [Cox] that the Coxeter groups varieties $\left\langle
p,q,r\right\rangle _{s}$ exhaust the topological types of \emph{resolvable
singularites}. This is just one aspect of

\textit{"the profound connections between the critical points of functions,
quivers, caustics, wavefronts, regular polyhedra,... and the theory of groups
generated by reflections"} [Arnold 2 ]. The profound connections\textit{\ }%
that are important here are

\ \qquad\qquad\qquad\qquad\qquad\qquad\qquad\qquad\qquad\ \ \ \ 

\begin{enumerate}
\item $L$ or $R$ multiplication by a \textit{spacelike} C vector gives a $L$-
or $R$-\emph{helicity twist} about axis $\mathbf{\hat{\ell}}$ or
$\mathbf{\hat{r}}$ by angle $\frac{\lambda}{2}$or $\frac{\rho}{2}:$
\end{enumerate}

$\ \ \ \ \ell^{\prime}=\mathbf{L}\left(  \lambda\right)  \ell\equiv\exp\left(
\frac{i\lambda}{2}\mathbf{\hat{\ell}}\cdot\mathbf{\sigma}\right)  \ell
;$\ \ $\overline{r}^{\prime}=\bar{r}\mathbf{\bar{R}}(\rho)\equiv r\exp\left(
\frac{i\rho}{2}\mathbf{\hat{r}}\cdot\mathbf{\sigma}\right)  \Rightarrow
q^{\prime}=\left(  \ell\otimes\bar{r}\right)  ^{\prime}=L\left(  \ell
\otimes\bar{r}\right)  \bar{R}$ $.$

\qquad\qquad\qquad\ \ \ 

$\ \ \ L$ or $R$ action generated by a \textit{null }C vector gives an
additional $U(1)$ twist :

$\ \ \ \ \ \ell^{\prime}=L\ell=\exp\frac{i\lambda}{2}\left[  \pm\sigma_{0}\pm
i\mathbf{\hat{\ell}}\cdot\mathbf{\sigma}\right]  $ ,$\ \ \ \ \bar{r}^{\prime
}=\bar{r}\bar{R}=\bar{r}\exp\frac{i\rho}{2}\left[  \pm\sigma_{0}\pm
i\mathbf{\hat{r}}\cdot\mathbf{\sigma}\right]  $ $\ \Rightarrow q^{\prime
}\equiv\left(  \ell\otimes\bar{r}\right)  ^{\prime}=L\left(  \ell\otimes
\bar{r}\right)  \bar{R}$ $.$

\begin{enumerate}
\item[2.] Conjugation by a spacelike C vector reflects a flag (a $\mathbf{3}$
vector, $\mathbf{q,}$ and its normal frame) in a mirror with unit normal
$\mathbf{a}$ (see appendix)$\mathbf{:}$ \ \ \ \ \ \ \ \ \ \ \ 
\end{enumerate}

$\ \ \ \ \ \ \ \ \ \mathbf{q}^{\prime}=-\mathbf{aqa}^{-1}=[i\mathbf{a]q[}%
i\mathbf{a}^{-1}]=[i\mathbf{a](}l\otimes r)\mathbf{[}i\mathbf{a}%
^{-1}]\Rightarrow l^{\prime}=[i\mathbf{a];}$ $r^{\prime}=r\mathbf{[}%
i\mathbf{a}^{-1}].$

An ordinary (period -2) reflection reverses the flag $L\leftrightarrows R,$
preserves function values, but reverses differentials, creating a
\textit{singularity} on the mirror plane. A domain $B_{3}$ \ bounded by
mirrors (like a laser cavity) becomes a \textit{resonator}: it traps waves at
its fundamental frequency or its \ harmonics to create a standing wave.\ 

\begin{enumerate}
\item[3.] Reflections in mirror planes $P_{\perp}$ and $Q_{\perp}$ that
intersect at dihedral angle $\frac{\theta}{2}$ give a \textit{rotation} by
$\ \theta$ around the (spacelike) axis. $\mathbf{a}=P_{\perp}\cap Q_{\perp}
$:$\ q^{\substack{^{\prime} \\^{\prime}}}=\ell^{\prime}\otimes\bar{r}^{\prime
}=L\left(  \ell\otimes\bar{r}\right)  L^{-1};$ $\ \ L=\exp\left(
\frac{i\lambda}{2}\mathbf{\hat{\ell}}\cdot\mathbf{\sigma}\right)  .$

\item[4.] $L$ or $R$ action by $\ $the complex Clifford ($\mathfrak{%
\mathbb{C}
}C$) vector, $\exp\frac{i\theta}{p}\left[  \pm\sigma_{0}\pm i\mathbf{\hat
{\ell}}\cdot\mathbf{\sigma}\right]  ,$ gives a \emph{period-}$p$
reflection.\ It takes $p$ repeated reflections to close a spatial cycle; this
first happens for $\theta=\pi,$ making an\ image with \emph{dihedral}
symmetry, $D_{p}$. Parallel mirrors a distance $\frac{\bigtriangleup}{2}$
apart generate \emph{translations} of $~\bigtriangleup$.

\item[6.] On $\ [M_{\#}]_{diag}\equiv\lbrack S_{l}(a_{\#})\times S_{3}%
(a_{\#)}]_{diag}$, multiple reflections in 3 planes that all intersect in
\emph{one point}\ form a 3D\textit{\ kaleidoscope }in spin space. Its image is
a \emph{discrete subgroup}, $R_{s}\subset U\left(  1\right)  \times SU\left(
2\right)  $, provided that the three dihedral angles,$\ \left(  \frac{\pi}%
{p},\frac{\pi}{q},\frac{\pi}{r}\right)  $ and the multiplicity, $s,$obey the
\emph{closure} constraint: to commute, all 4 arguments above must be multiples
of $\pi$ :
\end{enumerate}%

\begin{equation}%
\begin{array}
[c]{c}%
R_{P}\equiv\exp\left(  i\pi sp^{-1}P\right)  ,\quad R_{Q}\equiv\exp\left(
i\pi sq^{-1}Q\right)  ,\quad R_{R}\equiv\exp\left(  i\pi sr^{-1}R\right)
;\quad R_{S}\equiv\exp\left(  -i\pi s^{-1}\sigma_{0}\right)  ;\text{ \ }\\
\text{\ }R_{P}^{p}=R_{Q}^{q}=R_{R}^{r}=R_{S}^{-s}=R_{P}R_{Q}R_{R}R_{S}=\pm1\\
\Rightarrow s\left(  p^{-1}+q^{-1}+r^{-1}-1\right)  =n,
\end{array}
\end{equation}

where $n,$ and $(p,q,r)$ are all integers. The integer $s,$ a \textit{common
multiple} $\ s=\ C.M.\left(  p,q,r\right)  ,$ is the \textit{length of the
string of reflections} that reconstructs all of the images in the
representation $R_{s}$ of the \textit{Coxeter group, }$\left\langle
p,q,r\right\rangle _{s}:$ the $s-$ fold cover of the Rotation, Dihedral,
Tetrahedral, Octahedral, or Icosahedral group, $(A_{p},D_{p},T,E_{6},$
$E_{8}).$ The common multiple, $s,$ is called the \emph{multiplicity, }or
\textit{Coxeter number} [Coxeter].\textit{\ }

This brings us right to the main point:

\begin{center}
\textit{The mass- the 3-volume in spin space spanned by the string of }$s$
\textit{reflections }

$\left\langle p,q,r\right\rangle _{s},$ \textit{varies as the cube of the
string length:} $m\backsim(s/2)^{3}$ .\bigskip
\end{center}

More precisely, the rest energy of the configuration $\ \{\psi_{I},\psi^{I}%
\}$\textbf{- }i.e. its Noether charge under T translation, is

\begin{center}
$\bigskip\int_{M_{\#}}\{\partial%
\mathcal{L}%
_{g}\ /\partial_{T}\psi_{I}\}\ [\partial\psi^{I}\ /\partial T]_{0}$
$=\int_{M_{\#}}[\psi^{I}\mathbf{d}\psi_{I}]^{3}=(\frac{s}{2})^{3}$ $(\frac
{1}{2a_{\#}}$ $).$
\end{center}

\bigskip For a periodic solution to match the vacuum fields on the boundary
$\gamma_{3}=\partial B_{4},$ the \ \textit{frequency,} $\omega(s,n)$ inside a
particle's world tube must be a harmonic of the vacuum frequency; $\omega
_{0}=(2a_{\#})^{-1}.$ For the odd spin structure on $M_{\#}$ [Geroch], it
takes time $\Delta t=$ $2\pi a_{\#}$ for a lightlike phase front, $\theta
^{a}=const.,$ to circumnavigate a ray on a closed light cone,$\ \widehat{N}%
\in\lbrack S_{1}\times S_{3}(a_{\#})]_{diag}$ ; one circuit gives
$\Delta\theta^{0}=\Delta\theta^{j}=\pi,$ so $\psi(t+2\pi a_{\#})=\psi(t)$. A
solution of period $s$ contributes a mass increment inside its world tube of%

\begin{align}
m  & =(s/2)^{3}(2a_{\#})^{-1}.\\
&
\end{align}

The energy - the 3-volume in spin-space- is counted according to its
multiplicity, $s$: the number of \ spin-space sheets above the particle's
support. For $s=2,$ this is the\textit{\ }mass\textit{\ }of\textit{\ }an
\textit{electron}, governed by the\textit{\ massive Dirac }equations $(21)$
inside its world tube, $B_{4},$ of radius $a_{\#}$ $.$ For a free electron,
$e^{-}\equiv(\mathbf{l}\oplus\mathbf{r})\in\left\langle 2,2,2\right\rangle
_{2}$, all 3 dihedral angles are $\frac{\pi}{2}$. The 3 pairs of vacuum
spinors which trap the matter pair inside a 3-cube form opposing pairs of
\textit{corner-cube reflectors. }

As we decrease one of the dihedral angles, we get a 3- cycle at $\frac{\pi}%
{3}$; 3 sheets bounded by a \textit{tuck }caustic ref. [Arnold]. But the cycle
generated by \textit{both} reflections doesn't close up again until we reach
their least-common multiple (lcm), $2\cdot3=6.$ giving a 6-fold cover of the
reflection group: the Coxeter group, $<2,2,3>_{6},$ with multiplicity $6$. We
identify this as the \textit{muon; }and the next closed reflection cycle,
$<2,3,4>_{12}$as the \textit{tauon }. More generally,

\begin{center}
\textit{\ \ A massive lepton, meson, or hadron is composed of }$\ J=1,2,$%
\textit{or }$3$\textit{\ pairs of oppositely-propagating bispinor pairs,}

\textit{\ trapped inside a timelike world tube by reflections off interference
gratings with }$(4-J)$ \textit{vacuum pairs on its boundary. }
\end{center}

What is new here is that the reflection groups $<p,q,r>_{s}$of multiplicity
\ $s=(2,3,4,5,6,12,30)$ not only \textit{classify }the elementary particles,
but give their \textit{mass ratios }(table III)\textit{\ },%

\begin{equation}
\frac{m}{m_{e}}=(\frac{s}{2})^{3}.
\end{equation}

\textit{These agree with the observed mass ratios within a few percent
}(except for the $\pi$ mesons, which are off by $\backsim$ 25\% )\textit{\ }.

\begin{center}
\noindent$%
\begin{array}
[c]{c}%
\text{\textbf{Table III: Spin-}}J\text{ \textbf{Resonances:}}\\
\\
\text{Codimension-}J\text{ singularities in the }U\left(  1\right)  \times
SU\left(  2\right)  \text{ phase, };\\
\text{with wave fronts }x^{p}+y^{q}+z^{r}+T^{-s}=const\\
\text{the Brieskorn varieties. Each represents a closed cycle of Bragg
reflections of a }\\
\text{chiral pair of matter spinors, }\left(  \psi_{I},\psi^{I}\right)
,\text{off the interference gratings between }\\
\text{the remaining }\left(  J-1\right)  \text{ matter pairs and }\left(
4-J\right)  \text{ perturbed vacuum pairs.}\\%
\begin{array}
[c]{ccccccc}%
\text{Particle} &  &
\begin{array}
[c]{c}%
\text{Binary}\\
\text{Group}%
\end{array}
&  &
\begin{array}
[c]{c}%
\text{Coxeter}\\
\text{Numbers}%
\end{array}
&  & \frac{m}{m_{e}}\\
&  & H\subset\left[  SU\left(  2\right)  \right]  ^{J} &  & s:\left\langle
p,q,r\right\rangle _{s} &  &
\begin{array}
[c]{cc}%
\left(  \frac{s}{2}\right)  ^{3} & \text{obs}%
\end{array}
\\
e^{-} &  & D_{2} &  & \left\langle 2,2,2\right\rangle _{2} &  &
\begin{array}
[c]{cc}%
1 & 1
\end{array}
\\
&  & D_{p} &  & \left\langle 2,2,p\right\rangle _{p} &  & \\
&  & T &  & \left\langle 2,3,3\right\rangle _{6} &  & \\
\mu^{-} &  & O &  & \left\langle 2,3,4\right\rangle _{12} &  &
\begin{array}
[c]{cc}%
216 & 207
\end{array}
\\
\tau^{-} &  & I &  & \left\langle 2,3,5\right\rangle _{30} &  &
\begin{array}
[c]{cc}%
3375 & 3478
\end{array}
\\
\pi^{-} &  & D_{3}\otimes\bar{D}_{4} &  & \left\langle 2,2,3\right\rangle
_{3}\otimes\left\langle 2,2,4\right\rangle _{4} & d\bar{u} &
\begin{array}
[c]{cc}%
216 & 275
\end{array}
\\
k^{-} &  & D_{4}\otimes\bar{D}_{5} &  & \left\langle 2,2,5\right\rangle
_{5}\otimes\left\langle 2,2,4\right\rangle _{4} & s\bar{u} &
\begin{array}
[c]{cc}%
1000 & 975
\end{array}
\\
D_{s}^{-} &  & D_{5}\otimes\bar{D}_{6} &  & \left\langle 2,2,5\right\rangle
_{5}\otimes\left\langle 2,2,6\right\rangle _{6} & s\bar{c} &
\begin{array}
[c]{cc}%
3375 & 3647
\end{array}
\\
n_{c} &  & D_{6}\otimes\bar{D}_{6} &  & \left\langle 2,2,6\right\rangle
_{6}\otimes\left\langle 2,2,6\right\rangle _{6} & c\bar{c} &
\begin{array}
[c]{cc}%
5832 & 5686
\end{array}
\\
p^{+} &  & D_{4}\otimes D_{4}\otimes D_{3} &  & \left\langle
2,2,4\right\rangle _{4}\otimes\left\langle 2,2,3\right\rangle _{3}%
\otimes\left\langle 2,2,4\right\rangle _{4} & \left[  u,d\right]  u &
\begin{array}
[c]{cc}%
1728 & 1836
\end{array}
\end{array}
\end{array}
$
\end{center}

\bigskip

In the quantum calculation (III) we sum over histories in "imaginary time", T:
all possible chains of null zigzags connecting the initial and final states [MC3].

Microscopically, it seem, the whole world, both outside and inside the world
tubes of massive particles, resolves into a network of light-like spinors, and
their scattering vertices: their multilinear interactions. \ 

\ 

\section{{\protect\LARGE Conclusions and Open Question}}

Spin Principle \textbf{Pl} says that the 8-spinor bundle, $\mathbf{8}%
$\textbf{,} is the physical reality; and that the action is just its volume in
spin space. Our spacetime 4- fold, $M,$ and the particle wave functions,
$\Psi,$ are horizontal and vertical projections of a minimal-surface in spin
space: the \textit{spinfoam. }The regular stratum, or \textit{vacuum},
$D^{o},$ can be combed parallel locally by path-dependent phase differentials,
$\mathbf{d\zeta}_{I}\ =\Psi^{I}\mathbf{d\Psi}_{I},$ by spin connections: the
vector potentials. Their spin curvatures, $\Psi^{I}\mathbf{dd\Psi}_{I}$ , are
the fields, If these carry a nontrivial flux (topological charge) over the
boundary, it \textit{must }enclose a singularity-at least, in the
\textit{projection}, $\pi:$ $\mathbf{8\rightarrow}M$ : a \textit{caustic.
}Caustics are characterized by their symmetry groups in spin space, and there
are only a few admissible types: the Coxeter groups, $\langle p,q,r\rangle
_{s}$.

In the continuum picture, their representations are the wave functions of
particles with definite spin. They look like the \textit{Brieskorn varieties}:
fibred knots, whose isophase contours and normal rays ("lines of force")
radiate and terminate on singularities. \ Their masses - i.e. their Noether
time- translation charges, are $m=(s/2)^{3},$ in natural units of \ $2a_{\#}%
{}^{-1}$; the mass of the electron $(s=2).$

In the discrete picture, a vertex where a $r$-chirality spinor reflects from a
Bragg mirror $l\otimes\overline{r}$ into an oppositely-propagating $l$-
chirality one is called a\textit{\ mass scattering: } $(l\otimes\overline{r} $
$)r\rightarrow l$ \textit{\ }[Penrose]. A \textit{null zigzag} is a
\textit{pair} of mass scatterings, $L\rightarrow R\rightarrow L;$%
\textit{\ }the discrete version\textit{\ }of a \textit{fold}. To close a cycle
of null zigzags, \textit{each} chiral component must return to its original
value. This happens only after a common multiple (c. m.) of the three binary
reflection degrees, $s=cm(p,q,r)$. But it takes only $\frac{s}{2}$ reflections
to restore a \textit{bispinor }state; $R_{\frac{s}{2}}:\left(  \ell\oplus
r\right)  \rightarrow\left(  r\oplus\ell\right)  ,$ for $\frac{s}{2}$ odd.

Why should the reflection groups -the same groups that
classify\textit{\ resolvable} \textit{singularities, regular polyhedra, Lie
algebras, quivers, frieze patterns, honeycombs,} \textit{crystals, and
caustics}- classify the \textit{elementary particles}? Because they all arise
from the generic structures of singularities in flows.

Like heat flow resolves into random walks, at the critical scale, $a_{\#},$
the \ $\mathbf{8-}$ spinor flow resolves into a microhistory of \textit{null
zigzags}. In each discrete history, the \textit{multiplicity}, $s$- the number
of null zig-zags it takes to close a cycle- must be a common multiple of the
reflection degrees $p,q,$and\ $r$. This results in an image in spin space like
that formed by light rays crisscrossing in a 3D kaleidoscope, with mirrors at
angles $\frac{\pi}{p},\frac{\pi}{q},\frac{\pi}{r}.$The nonlinear 8-spinor
dynamics has appeared above as multilinear mode-mode coupling in \textit{spin
space}. This discrete picture of spin rays and Bragg reflections is only a
skeleton of the quantum dynamics of the 8-spinor system. To get the quantum
corrections to the particle masses, we must sum over all possible histories of
spin rays and intermediate scattering vertices; just as we sum over random
walks to get the heat propagator. The "random walk" underlying the Dirac
system - the \textit{Dirac propagator} - is the\textit{\ sum over all null
zigzag histories} connecting the initial and final states [Feynman], [Penrose
l],[Ord ].

What is subtle and beautiful about this picture is

1) how self-consistent cycles of $J$ chiral pairs of matter waves and $(4-J)$
vacuum pairs "bootstrap" each other into existence as the radius passes
through $\gamma=1,$ where $T=a_{\#};$ the critical radius for the inflationary
phase transition (\textrm{III}).

2) How the $J-$dimensional critical modes that "crystalize out" at
$O(\gamma^{J+1}),$ \textit{program the multilinear} \textit{couplings }of
modes at the next shorter scale, much as a volume hologram couples input to
output waves. This results in the ramification of patterns at smaller and
smaller scales, much like the main sequence of wavenumber-doubling
\ bifurcations leading to turbulence.

Is this what we're seeing in the sequence of $\ l=(200,400,800...)$ modes in
the Cosmic Microwave Background near the time of decoupling; or in the
foam-like structure of incident $J=(1,2,3)-$ branes in the large-scale
distribution of galaxies?

Perhaps the regular background of vacuum spinors is the\textit{\ dark energy}-
the invisible Dirac sea, on which the wave functions of visible matter ride
like waves on the surface of the ocean.

3)Since the Dirac mass term is created by products of vacuum spinors, these
might be called \textit{dark matter}. This picture not only shows how the
"distant masses" endow particles with their rest masses, but closely
approximates the measured particle masses\textit{.}

4)To get the quantum corrections, we must sum over all null zigzag histories
connecting initial and final states. Since particles\textit{\ are }null
zigzags climbing up a timelike worldtube, these intermediate histories could
be interpreted as the creation and anhialation of "virtual" particles; but it
is easier to sum zigzags on a null lattice than creations and anhinhialations
of all possible virtual particles.

We do this sum over histories for a Friedman universe and derive an effective
potential for its dilation flow in the sequel [M. C. 3]. Since we are drifting
along in this flow, like rafters on a river, we don't see it directly.
Instead, we see the scenery on the banks marching by in a sequence we call
\textit{time}.

\bigskip

\section{\bigskip Appendix: From Spin Space to Spacetime}

Spinors live in the "square-root space" of Clifford (C) vectors, or
\textit{Clifford tetrads, }$q_{\alpha}=(l\otimes_{\alpha}r)\in C(R_{4}):$ the
spin-1 representations of displacements in the \textit{Clifford algebra}
[BDS]. Their history goes back to Hamilton, who discovered the key to
composing rotations in space: express each rotation as a product of two
\textit{reflections}, and conjugate the\ vector argument with the string of
reflection operators. Pauli reexpressed these as $Sl(2,C)$ matrices, which
decompose into tensor products $q_{\alpha}\in$ of column and row spinors: the
fundamental representations of spin isometries(see appendix).

While Cartan and Clifford [Cliff] were developing the geometric role of
spinors, Weyl discovered that a $l$ or $r-$chirality spinor by itself
represents the wave function of a left or right-helicity \textit{neutrino,
}and Dirac discovered that their direct sum, $(l\oplus r)=e,$ represents on
electron. Meanwhile, Van der Waerden showed how to build any matter or
geometric field from tensor products $(l\otimes r)^{J}$ of $l-$ and
$r-$chirality spinors; the fundamental representations of the spin isometry
group, or \textit{Einstein Group, }\textbf{E, }of rotations\textbf{,
}translations, and boosts extended by P (parity), and T(cosmic) time reversal.
Hamilton expressed reflection operators as \textit{quaternions.}
Pauliexpressed these as \textit{spin matrices }in a basis of $2\times2$
antiHermitian matrices, with an algebra isomorphic to the quaternion algebra.

\begin{center}
\ \ \ \ \ \ \ \ \ \ \ \ \ \ \ \ \ \ \ \ \ \ \ \ \ \ \ \ \ \ \ \ \ \ \ \ \ \ \ \ \ \ \ \ \ \ \ \ \ \ \ \ \ \ \ \ 

Allowing \textit{independent \ spin transformations }to act from the left and
right, $V^{\prime}=aVb,$ gives $spin$ $4,$ the spin representation of $SO(4).$
The reflection of a spacelike C vector in a mirror with normal C
vector\ $a\equiv a^{k}q_{k}$ is expressed by

$V=%
\begin{bmatrix}
x_{3}-ix_{4} & x_{1}+ix^{2}\\
x_{1}-ix_{2} & -x_{3}+ix_{4}%
\end{bmatrix}
:V^{\prime}=aVa=-aVa^{-1}$ ;$\ V\equiv V^{1}q_{1}+V^{2}q_{2}+V^{3}q_{3}\equiv
V^{j}q_{j}$ .\ 
\end{center}

A rotation, $r,$ is composed of two reflections $a$ and $b.$

\begin{center}
\ \ \ \ \ \ \ \ \ \ \ \ \ \ \ \ \ \ \ \ \ \ \ \ \ \ \ \ \ \ \ \ \ \ \ \ \ \ \ \ \ \ \ \ \ \ \ \ \ \ \ \ \ \ \ \ \ \ \ \ \ \ \ \ \ \ \ \ \ \ \ \ \ \ \ \ \ \ \ \ \ \ \ \ \ \ \ \ \ \ \ \ \ \ \ \ \ \ \ \ \ \ \ \ \ \ \ \ \ 1)

$V^{\prime}=baVab\equiv rVr^{-1};$ $r=-ba$,

$V^{\prime}=r_{2}r_{1}Vr_{1}^{-1}r_{2}^{-1}.$
\end{center}

\bigskip Cartan defined the reflection operators $a$ and $a^{-1}$ in 1) as
acting from the left and from the right on the \textit{left and} \textit{right
chirality spinors, }with\textit{\ }a basis of 2 \textit{lightlike }column
vectors $l\equiv\{l_{1}^{T},l_{2}^{T}\}$ and 2 lightlike row vectors
$r\equiv\{r^{1},r^{2}\}.$a $[$ $2\times2]$ Their \textit{dyads }form a basis
for\ all $2\times2$ spin matrices; in particular\ the position/velocity spin
matrix for a particle a local inertial frame:

\begin{center}
$V=l\otimes r$ $:l=[l_{1}^{T},l_{2}^{T}]^{T}\in%
\mathbb{C}
_{2}$ ; \ $r=$\ $\{r^{1},r^{2}\}.$
\end{center}

Allowing \textit{independent \ spin transformations }to act from the left and
right, $V^{\prime}=aVb,$ gives $spin$ $4,$ the spin representation of $SO(4)$
This includes L and R -helicity \textit{screw translations} . Adjoining a
dilation generator $q_{4}\equiv$ $\frac{1}{2}\sigma_{0}$ gives translations
on$SO(4)$ This includes L and R -helicity \textit{screw translations} .
Adjoining a dilation generator $q_{4}\equiv$ $\frac{1}{2}\sigma_{0}$ gives
translations on $S_{3}\times R_{+}=R_{4}\backslash0.$

Adjoining the $U(1)$ generator $q$ $_{0}\equiv$ $\frac{i}{2}\sigma_{0}$ gives
$Spin(1,3):$ translations and rotations on compactified Minkowsky space,

$M_{\#}\equiv$ $S_{1}\times S_{3}$. Complexifying the timelike generator to
$Q_{0}\equiv$ $q_{0}\oplus iq_{4}$ gives $Spin^{c}4,$ which covers both
dilations and rotations

But it is only by complexifying all 4 generators to $Q_{\alpha}\equiv
q_{\alpha}\oplus ip_{\alpha}\in%
\mathbb{C}
T^{\ast}M,$ that we include Lorenz boosts, giving $Spin$ $(4,%
\mathbb{C}
):$ the generalization of the Poincare' group to $M_{\#}$ [ref]. Allowing the
position and world velocity $q_{\alpha}(x)$ and $p_{\alpha}(x)$ to vary
locally on $M_{\#}$ paints a \textit{dilation-boost flow}, $Spin$ $(4,%
\mathbb{C}
)_{loc}$ on a curved spacetime, $M;.$ $S_{3}\times R_{+}=R_{4}\backslash0:$
Wheeler's "lumpy potato".

Tensor products $\ q_{A}^{B}\equiv l_{A}\otimes r^{B}$ of 2 opposite-chirality
spinors make null spin vectors: photons. The \textit{null tetrads
}are\textit{\ }"vacuum photons", of helicity $\pm$1:%

\begin{equation}
q_{\upharpoonright}\equiv l_{+}\otimes r^{-};q_{\downarrow}\equiv l_{-}\otimes
r,\ q_{+}\equiv l_{+}\otimes r^{+};q_{-}\equiv l_{-}\otimes r^{-},
\end{equation}
Spin-1 sums of null tetrads make the \textit{Clifford tetrads }of a moving
Clifford algebra (C) frame:%

\begin{equation}%
\begin{array}
[c]{c}%
q_{a}\equiv\ell\otimes_{\alpha}r\in CT\mathbb{M}:\ \ \ \ q_{o}\equiv
(q_{\upharpoonright}-q_{\downarrow)};\ \ q_{1}\equiv(q_{+}+q_{-});\\
q_{2}\equiv-i(q_{+}-q_{-});\ \ q_{3}\equiv(q_{\upharpoonright}%
+q_{\downharpoonright}).
\end{array}
\end{equation}

The Clifford tetrads $q_{a}$ are identified with the basis vectors,
$e_{\alpha},$ of a spacetime frame via the\textit{\ }infinitesimal\textit{\ }%
form of the \textit{spin map, }$S:$ the canonical isomorphism of
\textit{compactified Minkowsky space},\textit{\ \ }$\mathbb{M}_{\#}%
=S_{1}\times S_{3}(a_{\#}),$to the compact Lie group, $U(1)\otimes SU(2):$

\begin{center}
$S\equiv\exp(\frac{i}{2a_{\#}})x^{\alpha}\sigma_{\alpha}:$ $\mathbb{M}%
_{\#}=S_{1}\times S_{3}(a_{\#})\rightarrow U(1)\otimes SU(2)\equiv g;$
\end{center}%

\begin{equation}
S_{\ast}\equiv g^{-1}\mathbf{d}g=(\frac{i}{2a_{\#}})q_{\alpha}e^{\alpha
}:e_{\beta}\rightarrow(\frac{i}{2a_{\#}})q_{\beta},\quad
\end{equation}

the Maurer-Cartan 1 form, valued in the Lie Algebra $u(1)\oplus su(2),$
generated by the Pauli spin matrrices $\sigma_{j},$ along with $\sigma
_{0}=\mathbf{1,}$ the 2x2 identity matrix. Here $\mathbf{d}\equiv e^{\alpha
}\left(  x\right)  \partial_{\alpha}\left(  x\right)  $ is the generalized
(possibly path-dependent) exterior differential operator; the $e_{\alpha}$ are
a moving frame of spacetime basis vectors, and the $q_{\alpha}(x)$ $\equiv
g\sigma_{\alpha}g^{-1}$ the isomorphic\textit{\ moving frame} \textit{in the
Clifford algebra. }The pullback of the spin map is the\textit{\ Dirac
operator, }$D=S^{\ast}:-i(2a_{\#})e^{\beta}\hookleftarrow q^{\beta}\equiv
\ell\otimes^{\beta}r,$ which assigns a spacetime increment to chiral pairs of
null spinors, \ $\overline{D}l=0;$ $Dr=0$ (3), where $\overline{D}%
\equiv\overline{q}^{\alpha}\partial_{\alpha}:$ $\overline{q}^{\alpha}%
\equiv(q_{0,}-q_{j}).$

More generally, in moving frames in spin space and spacetime, the spin map reads

\begin{center}
$%
\begin{array}
[c]{c}%
S_{\ast}(x)\equiv\mathbf{d\zeta(x)\equiv}[\partial_{\alpha}\zeta^{\beta
}](x)q_{\beta}(x)e^{\alpha}(x):e_{\alpha}(x)\rightarrow\left[  \partial
_{\alpha}\zeta^{\beta}\right]  q_{\beta}(x).
\end{array}
$
\end{center}

Its Jacobean determinant, $\left\vert \mathbf{d}\zeta\right\vert (x),$ is the
$4$- volume element in spin space; at a singular (critical) point, $x_{c} $
,$\ \left\vert \mathbf{d}\zeta\right\vert (x_{c)}=0$ . We use $\ \left\vert
\mathbf{d}\zeta\right\vert $ as our Lagrangian density.

\bigskip

\end{document}